\documentclass[%
twocolumn,
prl,
groupedaddress,
aps,
10pt
]{revtex4-2}

\usepackage{amsmath,amssymb,amsthm}

\usepackage{graphicx}
\usepackage[dvipsnames]{xcolor}
\usepackage{xspace}
\usepackage{upgreek}
\usepackage{mathtools}
\usepackage{newtxtext}
\usepackage{newtxmath}
\usepackage{braket}

\usepackage[colorlinks=true, 
urlcolor=blue, 
linkcolor=blue, 
citecolor=blue, 
hyperindex=true, 
linktocpage=true]{hyperref}

\newcommand{\hilb}{\mathcal{H}}
\newcommand{\Renyi}{R\'{e}nyi\xspace}

\DeclareMathOperator{\tr}{tr}
\newcommand{\ketbra}[2]{\ket{#1}{\bra{#2}}}

\newcommand{\ee}{\mathrm{e}}

\newtheorem{thm}{Theorem}
\newtheorem{res}[thm]{Result}

\makeatletter
\renewcommand{\p@subsection}{}
\renewcommand{\p@subsubsection}{}
\makeatother

\begin{document}

\title{Quantum \Renyi-Jarzynski Equality}

\author{Benjamin Bobell}
\email{ben.bobell@colorado.edu}
\affiliation{Department of Physics and Center for Theory of Quantum Matter, University of Colorado Boulder, Boulder, Colorado 80309 USA}

\author{Mert Okyay}
\affiliation{Department of Physics and Center for Theory of Quantum Matter, University of Colorado Boulder, Boulder, Colorado 80309 USA}

\author{Rahul Nandkishore}
\affiliation{Department of Physics and Center for Theory of Quantum Matter, University of Colorado Boulder, Boulder, Colorado 80309 USA}

\begin{abstract}
The Jarzynski equality %(JE) 
provides a strict link between nonequilibrium work and equilibrium free energy changes. Its typical quantum formulations, however, rely on measurement protocols that destroy coherence. In this Letter, we use the resource-theoretic approach to derive a non-destructive quantum Jarzynski equality conditioned on the outcomes of an arbitrary bath observable. This yields the \Renyi-Jarzynski equality, which quantifies a finite bath's drift from equilibrium under a non-adiabatic drive via the \Renyi $k$-divergence. We further demonstrate that the \Renyi-Jarzynski equality provides a tunable cost function for quantum optimal control problems where minimizing bath drift is desired, such as state preparation and gate design, enabling the minimization of cross-talk in finite quantum systems. Our toy model exhibits a transition between competing minima for some critical value of $k$, illustrating how the \Renyi order tunes sensitivity to different regions of a bath distribution. Strikingly, when drive parameters vary across bath energy levels, minimizing bath drift requires generating system-bath entanglement. 
\end{abstract}

\maketitle

\paragraph*
{Introduction.}Fluctuation theorems represent a profound milestone in statistical mechanics, providing exact mathematical relationships between nonequilibrium and equilibrium quantities. Perhaps the most celebrated example is the Jarzynski equality (JE)~\cite{OriginalJE}, which established that for a classical system satisfying micro-reversibility driven arbitrarily far from equilibrium, the average of the exponential of the work $W$ done on the system is strictly related to the equilibrium free energy difference $\Delta F_\mathcal{S}$ between the initial and final thermal states of a system $\mathcal{S}$:
\begin{equation}
    \mathbb{E}[\ee^{-\beta W}] = \ee^{-\beta \Delta F_\mathcal{S}}\, .
\end{equation}
Since Jarzynski's initial paper, a plethora of theoretical generalizations have been developed~\cite{Crooks_1999, jarzynski2000hamiltonian,tasaki2000jarzynskirelationsquantumsystems, kurchan2001quantumfluctuationtheorem, Seifert_2005, differentialfluctuationtheorem, Talkner_2009, Esposito_2009, Sagawa_2010, Morikuni_2011, CampisiTalknerRev, Funo_2013, Wei_2017}; for example, fully quantum formulations of the JE were rigorously established in Refs.~\cite{Aberg_2018, Alhambra_2016} using resource-theoretic approaches. JE enables the extraction of equilibrium properties from nonequilibrium trajectories in nanoscale devices \cite{singlemolecule2014,Spectroscopy, An_2014, Koski_2013}, small-scale biological systems \cite{testJE,PhysRevLett.99.068101,PREINER2007930,doi:10.1073/pnas.071034098, Collin_2005}, and various machine learning models \cite{neal2001annealed, diffusionmodelpaper,Rotskoff_2019,wu2020stochastic, Yasuda_2022,Caselle_2022, cuin2026learning}.

While quantum fluctuation theorems (QFTs) have been generalized beyond idealized thermal baths to account for arbitrary environments~\cite{Arbitraryopen,Manzano_2018}, these formulations predominantly rely on the two-point measurement (TPM) protocol to evaluate work. The TPM protocol, however, inherently destroys coherence by projecting the system into its energy eigenbasis. To our knowledge, a fully quantum JE that simultaneously captures the nonequilibrium dynamics of an arbitrary bath while circumventing the destructive nature of the TPM scheme has not been established. In this Letter, we derive a fully quantum, arbitrary-bath fluctuation theorem within the resource-theoretic framework \cite{Alhambra_2016, Masanes_2017}, conditioned on the measurement outcomes of an arbitrary bath observable $\mathcal{O}_\mathcal{B}$. For quasi-classical trajectories, this yields what we dub the \Renyi-Jarzynski equality (RJE): 
\begin{equation}
     \mathbb{E}_{P_F(b)}\big[\mathbb{E}_{P_F(w|b)}[\ee^{-\beta W}]^k\big] = 
     \ee^{(k-1) \mathrm{D}_k(P_0||P_F)-k\beta \Delta F_\mathcal{S} },
\end{equation}
where $\Delta F_{\mathcal{S}}$ is the change in equilibrium free energy of the system from the driving protocol, $P_0(b)$ and $P_F(b)$ represent the initial and final probability distributions over the measurement outcomes of $\mathcal{O}_\mathcal{B}$, respectively, and $\mathrm{D}_k(P_0||P_F)$ is the Rényi $k$-divergence. This equality represents an exact thermodynamic quantification of the bath's drift from its initial equilibrium state induced by a non-adiabatic drive. By utilizing the resource-theoretic framework, our approach avoids the projective energy measurements central to the TPM scheme. Instead, the only preferred basis is that of the chosen bath observable $\mathcal{O}_{\mathcal{B}}$, inherently preserving the coherence required for truly quantum operations.

Remarkably, the RJE can be expressed entirely in terms of an expectation value over a $k$-fold projected ensemble \cite{Goldstein_2015, Cotler_2023}. Specifically, the equality takes the operational form
\begin{multline}
    \tr_{\mathcal{W}^{\otimes k}}\left((\ee^{\beta H_{\mathcal{W}}})^{\otimes k}\rho_{\mathcal{W}}^{(k)}\right) \\
    = (K_{\mathcal{W}})^k \ee^{(k-1)  \mathrm{D}_k(P_0||P_F)-k\beta \Delta F_\mathcal{S}}, 
\end{multline}
where $\rho_{\mathcal{W}}^{(k)}=\tr_{\mathcal{S}^{\otimes k}}(\rho_{\mathcal{S}\mathcal{W}}^{(k)})$ denotes the weight marginal of the $k$-th moment of the projected system-weight ensemble after the drive, and $K_\mathcal{W}$ is a normalization dependent on the initial state of the weight. Our result recasts the purely information-theoretic \Renyi divergence as the expectation value of a $k$-copy work observable. Evaluating this divergence requires no tomography, only conditional work statistics. Sweeping $k$ then determines the entire family $\{\mathrm{D}_k\}$, targeting specific regions of the bath distribution.

To demonstrate our RJE's utility, we present an illustrative application for quantum optimal control, encompassing tasks such as state preparation and gate implementation. Whenever preserving an auxiliary system or finite environment during a driving protocol is desired, the RJE provides a physically motivated, tunable cost function to minimize the bath drift that is directly measurable via the nonequilibrium work distribution. Crucially, we find that entanglement generation between the system and bath may be necessary to minimize bath drift.

% {\bf Setup:}
\paragraph*{Setup.}To avoid the challenges facing various notions of work in quantum mechanics, we adopt the resource-theoretic formalism of thermodynamics \cite{Alhambra_2016,Masanes_2017}. This framework bypasses destructive measurements by appending an ideal 
%\mert{ideal means?} \ben{Ideal weight has an infinite spectrum}
weight (e.g. a free particle) to our system such that the work is defined by translations of this weight. 

We consider a tripartite setup with system $\mathcal{S}$, bath $\mathcal{B}$, and weight $\mathcal{W}$, with respective Hilbert spaces $\mathcal{H}_\mathcal{S}$, $\mathcal{H}_\mathcal{B}$, and $\mathcal{H}_\mathcal{W}$. Each subsystem carries a Hamiltonian $H_{\mathcal{S}}$, $H_{\mathcal{B}}$, and $H_{\mathcal{W}}$; we fix $H_{\mathcal{W}} = x_{\mathcal{W}}$, the position operator of the weight. All dynamics are implemented by \emph{thermodynamic operations} on the joint system $\mathcal{S}\mathcal{B}\mathcal{W}$, which are energy-conserving unitaries $V$ acting on $\mathcal{H}_{\mathcal{S}}\otimes \mathcal{H}_{\mathcal{B}}\otimes \mathcal{H}_{\mathcal{W}}$ that commute with translations on $\mathcal{W}$, i.e., we assume $[V, p_{\mathcal{W}}] = 0$.

To model a protocol where the system Hamiltonian is driven from $H_{\mathcal{S}}$ to 
$H_{\mathcal{S}}'$, we introduce an auxiliary qubit $\mathcal{L}$. The global Hamiltonian is
\begin{equation}
    H=H_{\mathcal{S}} \otimes |0\rangle \langle 0|_{\mathcal{L}} +H_{\mathcal{S}}' \otimes {|1\rangle\langle1|_{\mathcal{L}}} +H_{\mathcal{B}} +H_{\mathcal{W}}\, .
\end{equation}
Initializing the qubit in the state $\rho_{\mathcal{L}}=\ketbra{0}{0}_{\mathcal{L}} $, the driving protocol is implemented by a thermodynamic operation $V$. For an arbitrary state $\rho_{\mathcal{S}\mathcal{B}\mathcal{W}}$, 
\begin{equation}
    V(\rho_{\mathcal{S}\mathcal{B}\mathcal{W}}\otimes \ketbra{0}{0}_{\mathcal{L}})V^{\dagger} = \rho'_{\mathcal{S}\mathcal{B}\mathcal{W}}\otimes \ketbra{1}{1}_{\mathcal{L}}. 
\end{equation}
Such a $V$ that flips the qubit can be decomposed as
\begin{equation}
    V= U\otimes \ketbra{1}{0}_{\mathcal{L}} +  \tilde{U}\otimes \ketbra{0}{1} 
   _{\mathcal{L}},
\end{equation}
for some unitaries $U$ and $\tilde{U}$ that act only on $\mathcal{S}\mathcal{B}\mathcal{W}$.  For convenience, we drop the subscripts $\mathcal{S}\mathcal{B}\mathcal{W}$ and let $U$ always denote a unitary on $\mathcal{S}\mathcal{B}\mathcal{W}$, while $V$ acts on $\mathcal{S}\mathcal{L}\mathcal{B}\mathcal{W}$. While $U$ does not commute with the Hamiltonian $H_{\mathcal{S}\mathcal{B}\mathcal{W}}\coloneq H_{\mathcal{S}}+H_{\mathcal{B}}+H_{\mathcal{W}}$, energy conservation $[V,H]=0$ enforces
\begin{equation}\label{eq:energy_cons}
    U(H_{\mathcal{S}}+H_{\mathcal{B}}+H_{\mathcal{W}})U^{\dagger} = H_{\mathcal{S}}'+H_{\mathcal{B}}+H_{\mathcal{W}}.
\end{equation}
Then, evolution of $\mathcal{S}\mathcal{B}\mathcal{W}$ is $\Gamma_{\mathcal{S}\mathcal{B}\mathcal{W}}(\rho_{\mathcal{S}\mathcal{B}\mathcal{W}}) \coloneq U\rho_{\mathcal{S}\mathcal{B}\mathcal{W}}U^{\dagger}.$

\paragraph*{Single-copy fluctuation theorem.}We now consider our initial system and bath to be at equilibrium, $\rho_{\mathcal{S}(\mathcal{B})}=\ee^{-\beta H_{\mathcal{S}(\mathcal{B})}}/Z_{\mathcal{S}(\mathcal{B})}$, where $\beta$ denotes the \textit{initial} inverse temperature of the system and bath. After evolving this to $\rho_{\mathcal{S}\mathcal{B}\mathcal{W}}' = \Gamma_{\mathcal{S}\mathcal{B}\mathcal{W}}(\rho_{\mathcal{S}\mathcal{B}\mathcal{W}})$, we measure some observable $\mathcal{O}_{\mathcal{B}}$ that acts \textit{only} on the bath $\mathcal{B}$ with measurement outcomes $\{o^b_\mathcal{B}\}_b$ and probabilities $\{P_F(b)\}_b$, updating $\rho_{\mathcal{S}\mathcal{B}\mathcal{W}}'\to (\Pi_b^{\mathcal{B}} \rho_{\mathcal{S}\mathcal{B}\mathcal{W}}'\Pi_b^{\mathcal{B}})/P_F(b)$, where $\Pi_b^{\mathcal{B}}$ is the projector for measurement outcome $o_\mathcal{B}^b$ and $P_F(b) =\smash{\tr (\Pi_b^{\mathcal{B}} \rho_{\mathcal{S}\mathcal{B}\mathcal{W}}'\Pi_b^{\mathcal{B}}})$. Finally, we trace out the bath to leave a conditional state on $\mathcal{S}\mathcal{W}$. This entire process is captured by the completely positive (CP) map
\begin{equation}\label{eq:CP_phi}
    \Phi_b(\rho_{\mathcal{S}\mathcal{W}}) = \tr_{\mathcal{B}}\left[\frac{\Pi^{\mathcal{B}}_b}{P_F(b)}\Gamma_{\mathcal{S}\mathcal{B}\mathcal{W}}\left(\rho_{\mathcal{S}\mathcal{W}}\otimes\frac{\ee^{-\beta H_{\mathcal{B}}}}{Z_{\mathcal{B}}}\right)\right].
\end{equation}

While this map is not trace-preserving due to the measurement $\Pi_b^{\mathcal{B}}$, the normalization factor $P_F(b)$ ensures it preserves the unit trace of the initial state.
Furthermore, as with any measurement, the bath measurement can be implemented unitarily via Stinespring dilation~\cite{stinespring1955positive}, appending dilated degrees of freedom to the system $\mathcal{S}$ without loss of generality. We further define the following CP ``twirl" on any subspace of $\hilb$, 
\begin{equation}\label{eq:twirl}
    \mathcal{J}_{\mathcal{G}}(\rho) =\ee^{\beta \mathcal{G}/2}\rho \ee^{\beta \mathcal{G}/2} \, , 
\end{equation}
with inverse $\mathcal{J}_{-\mathcal{G}}(\rho)$;
it reweighs the configurations with respect to $\mathcal{G}$, analogous to the $\exp(\beta W)$ term in the classical JE.

Using Eq.~\eqref{eq:twirl}, we can state our first fluctuation theorem (with channel composition implied):
\begin{res} 
\label{res:gibbsstochasticfine}
   Given the map
   \begin{equation}
       \Omega_{b}(\rho_{\mathcal{S}}\otimes\rho_{\mathcal{W}}) \coloneq \tr_{\mathcal{W}}\left(\mathcal{J}_{H_{\mathcal{S}}'+H_{\mathcal{W}}}\Phi_b\mathcal{J}_{-H_{\mathcal{W}}}(\rho_{\mathcal{S}}\otimes \rho_{\mathcal{W}})\right)\, , 
       \label{eq:gibbschannel}
   \end{equation}
   then
   \begin{equation}
       \Omega_{b}\left(\frac{\ee^{-\beta H_{\mathcal{S}}}}{Z_{\mathcal{S}}}\otimes\rho_{\mathcal{W}}\right) = \frac{P_0(b)}{Z_{\mathcal{S}}P_F(b)} \mathbb{I}_{\mathcal{S}},
   \end{equation}
   where $P_0(b)= \tr_{\mathcal{B}}\left( \Pi_b^{\mathcal{B}} \ee^{-\beta H_{\mathcal{B}}}\Pi_b^{\mathcal{B}} \right)/Z_{\mathcal{B}}$.
\end{res} 

Reading the composition in Eq.~\eqref{eq:gibbschannel} from right to left provides its physical intuition: the initial map $\mathcal{J}_{-H_{\mathcal{W}}}$ evaluates the exponential of the weight's initial energy, $\Phi_b$ executes the driving protocol and measurement, and the final map $\mathcal{J}_{H_{\mathcal{S}}'+H_{\mathcal{W}}}$ evaluates the final exponential energy and work. Symmetrically sandwiching the drive between these maps is necessary to preserve the hermiticity and positivity of the operators, which generally do not commute in the quantum regime.

We derive Result~\ref{res:gibbsstochasticfine} in the Appendix and note that it is a straightforward extension of Ref.~\cite{Alhambra_2016}. In Result \ref{res:gibbsstochasticfine}, we denote the distribution $ P_0(b)$ because it provides the probability of obtaining outcome $b$ \textit{before} the driving protocol occurs. Assuming the initial product state $\rho_{\mathcal{S}\mathcal{W}}\otimes \ee^{-\beta H_{\mathcal{B}}}/Z_{\mathcal{B}}$, the probability of obtaining outcome $b$ after measuring $\Pi_b^{\mathcal{B}}$ is
\begin{equation}
    \tr\left(\Pi_b^{\mathcal{B}} \rho_{\mathcal{S}\mathcal{W}}\otimes \frac{\ee^{-\beta H_{\mathcal{B}}}}{Z_{\mathcal{B}}} \right) =\tr_{\mathcal{B}}\left( \Pi_b^{\mathcal{B}} \frac{\ee^{-\beta H_{\mathcal
    B}}}{Z_{\mathcal{B}}}\right),
\end{equation}
which recovers the definition of $P_0(b)$. From Result \ref{res:gibbsstochasticfine}, we can derive a bath-measurement Jarzynski equality by applying $\mathcal{J}_{-H_{\mathcal{S}}'}$ to both sides and taking the trace over $\mathcal{S}$, which yields the following result.
\begin{res}[Bath Measurement Quantum Jarzynski Equality]
\label{QJEfine}
    Given the map
    \begin{equation}
    \begin{aligned}
        \mathcal{Y}_b(\rho_{\mathcal{S}}\otimes\rho_{\mathcal{W}}) \coloneq& \tr_{\mathcal{S}\mathcal{W}}\left(\mathcal{J}_{H_{\mathcal{W}}}\Phi_b\mathcal{J}_{-H_{\mathcal{W}}}(\rho_{\mathcal{S}}\otimes \rho_{\mathcal{W}})\right) \\
        \equiv &\tr_{\mathcal{S}}\left(\mathcal{J}_{-H_{\mathcal{S}}'}\Omega_{b}(\rho_{\mathcal{S}}\otimes\rho_{\mathcal{W}})\right)\, , 
    \end{aligned}
    \end{equation}
    it follows that
    \begin{equation}
         \mathcal{Y}_b\left(\frac{\ee^{-\beta H_{\mathcal{S}}}}{Z_{\mathcal{S}}}\otimes\rho_{\mathcal{W}}\right) = \ee^{-\beta \Delta F_{\mathcal{S}}}\frac{P_0(b)}{P_F(b)}. 
    \end{equation}
\end{res}
Here, we used that the free energy $F_{\mathcal{S}}=-\log(Z_\mathcal{S})/\beta$ to rewrite $Z_{\mathcal{S}}'/Z_{\mathcal{S}}=\ee^{-\beta \Delta F_{\mathcal{S}}}$. Although Result \ref{QJEfine} appears structurally distinct from the standard Jarzynski equality, taking a quasi-classical limit recovers a %familiar Jarzynski-like relation. 
more familiar form. Here, the ``limit" corresponds to the vanishing of quantum coherence, where the off-diagonal density matrix elements in the energy eigenbasis for $H_{\mathcal{S}}$ and $H_{\mathcal{S}}'$ are taken to zero.

This amounts to assuming that the channel $\Phi_b$ is diagonal in the energy eigenbasis.
\begin{equation}\label{eq:quasicl_phi}
    \Phi_b(\ketbra{s}{s}  \otimes \ketbra{0}{0}_{\mathcal{W}} ) = \sum_{s',w} P(s',w|s,b) \ketbra{s'}{s'} \otimes \ketbra{w}{w} ,
\end{equation}
where $\ket{s}$ and $\ket{s'}$ are energy eigenstates of $H_\mathcal{S}$ and $H_{\mathcal{S}}'$, respectively, and $\ket{w}$ is an energy eigenstate of $H_{\mathcal{W}}$. Further, $ P(s',w|s,b)$ is the probability of transitioning to energy eigenstates $\ket{s'},\ket{w}$ conditioned on starting in energy eigenstate $\ket{s}$ and yielding bath measurement outcome $b$. Because the work system is translationally invariant, we can say that $\rho_{\mathcal{W}}=\ketbra{0}{0}_{\mathcal{W}}$. Plugging into Result~\ref{QJEfine}, we get
\begin{res}
\label{res:quasiclassicalmeasurementjarz}
Under the quasi-classical approximation~\eqref{eq:quasicl_phi}, the conditional quasi-classical Jarzynski equality holds:
    \begin{equation}
         \mathbb{E}_{ P_F(w|b)}\left[ \ee^{\beta w}\right]= \frac{P_0(b) }{P_F(b) } \ee^{-\beta \Delta F_{\mathcal{S}} }.
    \end{equation}
\end{res} Result~\ref{res:quasiclassicalmeasurementjarz} may seem to have the wrong sign in the exponent,

but $w$ represents the energy change of the \textit{weight}, whose sign is strictly opposite to the work $W$ done \textit{on} $\mathcal{S}\mathcal{B}$. 

Furthermore, averaging over measurement outcomes reproduces the regular quasi-classical Jarzynski equality~\cite{OriginalJE, Alhambra_2016}, where
\begin{equation}
\begin{split}
    &\mathbb{E}[\ee^{\beta w}]=\mathbb{E}_{P_F(b)}\left[\mathbb{E}[\ee^{\beta w}]_{ P(w|b)}\right] \\&\quad = \sum_{b}P_F(b)\frac{P_0(b) Z_{\mathcal{S}}'}{P_F(b) Z_\mathcal{S}}
    =\ee^{-\beta \Delta F_\mathcal{S}}.
\end{split}
\end{equation}
Finally, we observe that taking the logarithm of both sides of Result~\ref{res:quasiclassicalmeasurementjarz}, averaging with respect to $P_F(b)$, applying Jensen's inequality and switching to work $w=-W$
yields a generalized second law~\cite{Esposito_2009}
\begin{equation}
\label{eq:klgeneralizedsecondlaw}
    \beta \mathbb{E}_{P_F(W)}[W]  \geq \mathrm{D}_{KL}(P_F||P_0) +\beta \Delta F_{\mathcal{S}},
\end{equation}
obtainable from Eq. 1 of~\cite{esposito2019} by using the non-negativity of the mutual information between system and bath.

\paragraph*{Bath-Measurement QFTs with multiple copies.}It is well known in quantum information that many copies of a system provide additional correlations and structure that can be exploited for advantage. Thus far, we have derived fluctuation theorems on a single copy of a system. In this section, we show that many-copy systems themselves also possess fluctuation theorems.

To extend the first-moment equalities with bath measurements we derived to higher moments, we first need to define the moments of interest, which are projected ensembles~\cite{Goldstein_2015,Cotler_2023}. If we possess a bipartite state $\rho_{\mathcal{A}\mathcal{B}}$, performing a projective measurement on $\mathcal{B}$ yields the conditional state $\rho_{\mathcal{A}|b}=\tr_\mathcal{B}(\Pi^\mathcal{B}_b\rho_{\mathcal{A}\mathcal{B}}\Pi^\mathcal{B}_b)/P(b)$, where $P(b)$ is the probability of outcome $b$. The projected $k$-th moment of this ensemble is defined as a density matrix on a $k$-fold cloned Hilbert space
\begin{equation}
    \rho_\mathcal{A}^{(k)}=\sum_b P(b) (\rho_{\mathcal{A}|b})^{\otimes k}.
\end{equation}
Our objective is to constrain these higher moments at the end of the driving protocol described in the previous section using the weight formalism, wherein $\mathcal{A} = \mathcal{S}\mathcal{W}$. As before, the channel $\Phi_b$~\eqref{eq:CP_phi} takes the initial system state $\rho_{\mathcal{S}\mathcal{W}}$, entangles it with the bath, performs a projective measurement on $\mathcal{B}$ yielding outcome $b$, and returns the conditional state $\rho_{\mathcal{S}\mathcal{W}|b}$, in terms of which~\eqref{eq:gibbschannel} can be rewritten as
\begin{equation}
\begin{split}
    &\Omega_b\left(\frac{\ee^{-\beta H_\mathcal{S}}}{Z_\mathcal{S}}\otimes\rho_\mathcal{W}\right) \equiv K_\mathcal{W}\text{Tr}_\mathcal{W}\left(\mathcal{J}_{H_\mathcal{S}'+H_\mathcal{W}}(\rho_{\mathcal{S}\mathcal{W}|b})\right),
\end{split}
\end{equation}
where $K_\mathcal{W} \coloneq \text{Tr}_\mathcal{W} (\ee^{-\beta H_\mathcal{W}}\rho_\mathcal{W})$. Since $\mathcal{J}_H$ is not trace preserving (TP), $K_\mathcal{W}$ ensures that the states inside $\Phi_b$ remain valid density matrices with $\tr \rho = 1$. Eq.~\eqref{eq:gibbschannel} can be copied $k$ times to define
\begin{multline}
    \label{eq:kgibbs} \Omega_b^{(k)}\left(\frac{\ee^{-\beta H_\mathcal{S}}}{Z_\mathcal{S}}\otimes \rho_\mathcal{W}\right)\coloneq\\
    K_\mathcal{W}^k \tr_{\mathcal{W}_1\otimes\cdots\otimes \mathcal{W}_k}\left(\mathcal{J}^{(k)}_{H_\mathcal{S}'+H_\mathcal{W}}\left(\rho_{\mathcal{S}\mathcal{W}|b}^{\otimes k}\right)\right),
\end{multline}
where we defined the $k$-fold twirl
\begin{equation}
    \mathcal{J}^{(k)}_{\mathcal{G}}(\rho) = \exp\left(\frac{\beta}{2}\mathcal{G}\right)^{\otimes k}\rho \exp\left(\frac{\beta}{2}\mathcal{G}\right)^{\otimes k},
\end{equation}
with its inverse denoted by $\mathcal{J}^{(k)}_{-\mathcal{G}}$. Since all the copies commute, Eq.~\eqref{eq:kgibbs} factorizes
\begin{equation}
\begin{aligned}
      \Omega_b^{(k)} &= \left(\tr_\mathcal{W}\left(K_\mathcal{W}\mathcal{J}_{H_\mathcal{S}'+H_\mathcal{W}}\left(\rho_{\mathcal{S}\mathcal{W}|b}\right)\right)\right)^{\otimes k} \\
      &=  \left(\Omega_b\left(\frac{\ee^{-\beta H_\mathcal{S}}}{Z_\mathcal{S}}\otimes\rho_\mathcal{W}\right) \right)^{\otimes k},
\end{aligned}
\end{equation}
where in the last step we recognized the term as Eq.~\eqref{eq:gibbschannel}. Using Result \ref{res:gibbsstochasticfine}, we can calculate $\Omega_b^{(k)}$
\begin{equation}
    \label{eq:gibbskresult}
    \Omega_b^{(k)}\left(\frac{\ee^{-\beta H_\mathcal{S}}}{Z_\mathcal{S}}\otimes \rho_\mathcal{W}\right) = \frac{P_0(b)^k}{Z_\mathcal{S}^kP_F(b)^k} \mathbb{I}_{\mathcal{S}_1\otimes\cdots \otimes \mathcal{S}_k}.
\end{equation}

We can similarly use Eq.~\eqref{eq:kgibbs} to define a $k$-fold Jarzynski map
\begin{equation}
    \begin{aligned}
    \mathcal{Y}_b^{(k)}\left(\cdot\right)\coloneq& K_\mathcal{W}^k \tr_{(\mathcal{S}\mathcal{W})_1\otimes\cdots\otimes (\mathcal{S}\mathcal{W})_k}\left(\mathcal{J}^{(k)}_{H_\mathcal{W}}\left(\rho_{\mathcal{S}\mathcal{W}|b}^{\otimes k}\right)\right)\\
    =&\tr_{\mathcal{S}_1\otimes \cdots \otimes \mathcal{S}_k}\left(\mathcal{J}^{-(k)}_{H_{\mathcal{S}}'}  \Omega_b^{(k)}\left(\cdot\right) \right).
\end{aligned}
\end{equation}
Applying Eq.~\eqref{eq:gibbskresult} yields the generalized equality
\begin{equation}
    \label{eq:kjarzynskiresult}
    \mathcal{Y}_b^{(k)}\left(\frac{\ee^{-\beta H_\mathcal{S}}}{Z_\mathcal{S}}\otimes \rho_\mathcal{W}\right) = \frac{P_0(b)^k}{P_F(b)^k} \ee^{-k\beta \Delta F_\mathcal{S}}.
\end{equation}
Eqs.~\eqref{eq:gibbskresult} and~\eqref{eq:kjarzynskiresult} allow us to construct maps that average over all copies. Because our maps are strictly linear, this provides direct fluctuation theorems for the $k$-th moments
\begin{align}
    \Omega^{(k)}(\cdot)&\coloneq\sum_{b}P_F(b)\Omega_b^{(k)}(\cdot) \notag\\
    &=  K_\mathcal{W}^k \tr_{\mathcal{W}_1\otimes\cdots\otimes \mathcal{W}_k}\left(\mathcal{J}^{(k)}_{H_\mathcal{S}'+H_\mathcal{W}}\left(\rho_{\mathcal{S}\mathcal{W}}^{(k)}\right)\right),  \\
    \mathcal{Y}^{(k)}\left(\cdot\right)&\coloneq\sum_{b}P_F(b)   \mathcal{Y}_b^{(k)}(\cdot) \notag \\
    &= K_\mathcal{W}^k \tr_{(\mathcal{S}\mathcal{W})_1\otimes\cdots\otimes (\mathcal{S}\mathcal{W})_k}\left(\mathcal{J}^{(k)}_{H_\mathcal{W}}\left(\rho_{\mathcal{S}\mathcal{W}}^{(k)}\right)\right).
\end{align}
Evaluating these averaged channels directly yields our main results
    \begin{multline}
        \label{eq:renyigibbs}
        \Omega^{(k)}\left(\frac{\ee^{-\beta H_\mathcal{S}}}{Z_\mathcal{S}}\otimes\rho_\mathcal{W}\right)=\frac{1}{Z_\mathcal{S}^k}\sum_{b} \frac{P_0(b)^k}{P_F(b)^{k-1}} \mathbb{I}_{\mathcal{S}_1\otimes\cdots \otimes \mathcal{S}_k} \\
        =Z_{\mathcal{S}}^{-k}\exp((k-1)\mathrm{D}_k(P_0||P_F)) \mathbb{I}_{\mathcal{S}_1\otimes\cdots \otimes \mathcal{S}_k}.
    \end{multline} 

\begin{res}[Quantum RJE]
    \label{res:renyiJE}
    \begin{multline}
       \mathcal{Y}^{(k)}\left(\frac{\ee^{-\beta H_\mathcal{S}}}{Z_\mathcal{S}}\otimes \rho_\mathcal{W}\right)=\ee^{-k\beta \Delta F_\mathcal{S}}\sum_{b} \frac{P_0(b)^k}{P_F(b)^{k-1}} \\
       =\ee^{-k\beta \Delta F_\mathcal{S}}\exp((k-1)\mathrm{D}_k(P_0||P_F)). 
    \end{multline} 
\end{res}
For both Eq.~\eqref{eq:renyigibbs} and Result~\ref{res:renyiJE}, we defined the \Renyi divergence of order $k$
\begin{equation}\label{eq:renyi_k_def}
    \mathrm{D}_k(P||Q)\coloneq \frac{1}{k-1} \log\left(\sum_{b}\frac{ P(b)^{k}}{Q(b)^{k-1}}\right),
\end{equation}
for outcome distributions $P$ and $Q$. The \Renyi divergence $\mathrm{D}_k(P||Q)$ provides a rigorous, information-theoretic quantification of the distinguishability between probability distributions across various moments \cite{minka2005divergence,li2016renyi}. Crucially, the order $k$ dictates the type of penalty in $Q$: a large $k$ penalizes $Q$ for underestimating $P$ (`mode-covering' behavior), while a smaller $k$ penalizes $Q$ for assigning probability to regions where $P$ is near zero (`mode-seeking' behavior). Thus, $k$ acts as a tunable parameter to control the tradeoff between various optimal distributions based on desired physical constraints. 

Taking the quasi-classical form of Eq.~\eqref{eq:kjarzynskiresult} (or equivalently, taking Result~\ref{res:quasiclassicalmeasurementjarz} to the $k$-th power and averaging over measurements), we get the quasi-classical version 
\begin{res}[Quasi-classical RJE]
    \begin{equation}
        \mathbb{E}_{P_F(b)}\left[\mathbb{E}[ \ee^{\beta w}]_{ P_F(w|b)}^k\right] = \ee^{-k\beta \Delta F_\mathcal{S}} \ee^{(k-1)\mathrm{D}_k(P_0||P_F)}.
    \end{equation}\label{quasiclassicalrenyik}
\end{res}
Applying Jensen's inequality (and using $w=-W$) produces a generalized second law incorporating the \Renyi divergence:
\begin{equation}
\label{eq:renyigeneralizedsecondlaw}
    \beta \mathbb{E}_{P_F(W)}[W] \geq \beta \Delta F_{\mathcal{S}}- \frac{k-1}{k}\mathrm{D}_{k} (P_0||P_F).
\end{equation}
Note that the status of $P_0$ and $P_F$ in Eq.~\eqref{eq:klgeneralizedsecondlaw} has swapped: $P_0$ is now the target distribution while $P_F$ is the reference distribution. 

Crucially, the above results place a strict thermodynamic constraint on the $k$-th moments of the projected ensemble of $\mathcal{SW}$. To our knowledge, Result~\ref{res:renyiJE} represents the first exact constraint placed on higher moments in fully non-equilibrium quantum thermodynamics. In quantum thermalization, the convergence of these higher moments to a universal maximum-entropy ensemble is dubbed \textit{deep thermalization} \cite{Cotler_2023, Mark_2024}. Since JE informs us about the first moment of the thermal ensemble, one might hope that our $k$-th moment constraint could therefore be used to probe deep thermalization of the system $\mathcal{S}$.  We do not believe, however, that this is the case: It is entirely possible that $\mathcal{S}$ undergoes deep thermalization while the joint $\mathcal{S}\mathcal{W}$ does not;  an arbitrary driving protocol may scramble information across $\mathcal{S}$ while leaving $\mathcal{W}$ with memory of its initial conditions. We leave a detailed investigation connecting fluctuation theorems to deep thermalization to future work.

\paragraph*{Application to Quantum Control.}\label{sec:quantumcontrol}
Result~\ref{res:renyiJE} provides a rigorous framework for quantifying the extent to which a driving process on a primary system $\mathcal{S}$ perturbs a coupled bath $\mathcal{B}$ out of equilibrium. This is particularly pronounced when $\mathcal{B}$ is a finite-size environment which interacts sufficiently strongly with $\mathcal{S}$, and the driving process operates at a finite, non-adiabatic rate. In other words, a large bath is barely disturbed by the drive, so $P_F\to P_0$ and the \Renyi divergence will vanish. We anticipate this framework will be particularly valuable for quantum optimal control and state preparation. We now present an illuminating application.

Consider a bipartite quantum system initialized in a state $\rho_{\mathcal{S}\mathcal{B}}=\rho_\mathcal{S}\otimes \rho_\mathcal{B}$ and suppose our goal is to execute some operation $\rho_{\mathcal{SB}}\to \rho_{\mathcal{SB}}'$ that keeps the final reduced state $\rho'_\mathcal{B}$ as similar as possible to $\rho_\mathcal{B}$. Even if we know the exact process to implement, this can be quite challenging due to cross-talk and noise on real hardware. Result~\ref{res:renyiJE} provides a rigorous metric to quantify this drift by mapping the distributions to effective Hamiltonians via $\rho_{\mathcal{S}(\mathcal{B})} \propto\ee^{-\beta H_{\mathcal{S}(\mathcal{B})}}$. Strictly speaking, Result~\ref{res:renyiJE} requires $\rho_\mathcal{S}$ and $\rho_\mathcal{B}$ to be full-rank mixed states, because they are both non-zero temperature thermal states. However, we can define approximately pure states via regularization; for example, if the target $\rho_\mathcal{S} =\ketbra{\uparrow}{\uparrow}$, we could define a Hamiltonian such that $\rho_\mathcal{S} = (1-\epsilon)\ketbra{\uparrow}{\uparrow} +\epsilon \ketbra{\downarrow}{\downarrow}$, for some small $\epsilon>0$. When designing optimal control sequences, the \Renyi $k$-divergence in Result~\ref{res:renyiJE} serves as a highly tunable loss function for the bath $\mathcal{B}$ that penalizes catastrophic tail-event errors much more aggressively than the standard Kullback-Leibler (KL) divergence.

Two distinct control problems arise. In the first, $H_{\mathcal{S}}'$ is chosen such that its thermal state is the desired target; the drive is judged by how close the final state is to that target thermal state and a quasi-static drive is optimal, returning the bath to equilibrium such that $\mathrm{D}_k(P_0||P_F)\to 0$. The \Renyi divergence thus quantifies the penalty incurred for a finite-rate drive of a finite bath. In the second, $H_{\mathcal{S}}\to H_{\mathcal{S}}'$ is an operation fixed in advance by hardware or by the gate to be implemented and the drive is optimized such that the final state approaches a predesignated target on $\mathcal{S}$ while perturbing the bath as little as possible. The target is generically not thermal with respect to $H_{\mathcal{S}}'$ and the bath may be unable to return to equilibrium under any admissible drive, in which case the \Renyi divergence will be strictly positive.

We present two closely-related two-qudit examples as a proof of principle for the second type of control problem. 

First, suppose the primary system $\mathcal{S}$ and bath $\mathcal{B}$ are each single qubits, interacting with an ideal weight $\mathcal{W}$, governed by the initial Hamiltonian  
\begin{equation}
    H_{\mathcal{S}\mathcal{B}\mathcal{W}}= \omega_\mathcal{S} \ketbra{1}{1} _\mathcal{S}+\omega_\mathcal{B}\ketbra{1}{1}_\mathcal{B}+\hat{x}_\mathcal{W}, 
\end{equation}
with $\mathcal{S}\mathcal{B}$ in a thermal state and unentangled with $\mathcal{W}$:
\begin{equation}
    \rho_{\mathcal{S}\mathcal{B}\mathcal{W}} = \frac{1}{Z_\mathcal{S}Z_\mathcal{B}}\ee^{-\beta (\omega_\mathcal{S} \ketbra{1}{1}_{\mathcal{S}}+\omega_{\mathcal{B}}\ketbra{1}{1}_\mathcal{B})}\otimes \ketbra{0}{0}_{\mathcal{W}}.
\end{equation}
We drive the system to change $\omega_{\mathcal{S}} \to \omega_\mathcal{S}'$ in the Hamiltonian $H_{\mathcal{S}}$; such protocols are of interest for quantum gate implementation or state preparation~\cite{DiCarlo_2009, Martinis_2014}. Because our Hamiltonian is diagonal in the $Z$ basis, we make measurements of $Z_\mathcal{B}$ after the drive, whose basis we denote $\ket{s}_\mathcal{S}\otimes \ket{b}_\mathcal{B}\equiv \ket{sb}$. Suppose we want the driving process to leave $\rho_{\mathcal{SB}}$ unscathed, 

however accidental entanglement is nevertheless generated when the qubits $\mathcal{S}\mathcal{B}$ differ: $U\ket{00} = \ket{00}$, $U\ket{11}=\ket{11}$, and
\begin{equation}
    U\begin{pmatrix} \ket{01}  \\ \ket{10} \end{pmatrix} = R(\theta)\begin{pmatrix}
        \ket{01} \\
        \ket{10}
    \end{pmatrix},
\end{equation}
for some $\theta$ that parameterizes the amount of entanglement generated and $R(\theta)\in \mathrm{SO}(2)$ is a rotation matrix
\begin{equation}\label{eq:rotMatrix_2}
   R(\theta)= \begin{pmatrix}
        \cos(\theta) & -\sin(\theta)\\
        \sin(\theta) &\hphantom{-}\cos(\theta) 
    \end{pmatrix} .
\end{equation}

The explicit form of $U$, including its necessary translation action on the weight to switch the Hamiltonian on $\mathcal{S}\mathcal{B}\mathcal{W}$, i.e. $U(H_\mathcal{S}+H_\mathcal{B}+H_\mathcal{W})U^{\dagger} = H_\mathcal{S}'+H_\mathcal{B}+H_\mathcal{W}$, is detailed in the Appendix. Note that, because the Hamiltonian has changed, the post-drive state is no longer thermal, even if it remains constant.  We now ask: what is the optimal drive (i.e. choice in $\theta$) that leaves the bath state as unchanged as possible? Answering this question for more complex models is practically relevant for mitigating crosstalk by ensuring that neighboring qubits remain unentangled \cite{Mundada_2019, ding2020systematic, lu2024camel}. Applying result~\ref{res:renyiJE} to minimize the \Renyi $k$-divergence implies that we must have $\theta = n\pi$ with $n \in \mathbb{Z}$, a fact that is intuitively clear from the form of~\eqref{eq:rotMatrix_2}. However, the true utility of our approach arises in cases where the structure of the entanglement is less analytically tractable. In such cases, the minima of \Renyi-$k$~\eqref{eq:renyi_k_def} can still be found experimentally or numerically by computing the non-equilibrium conditional expectation value over exponentiated work.

\begin{figure}[t]
    \centering
    \includegraphics[width=\columnwidth]{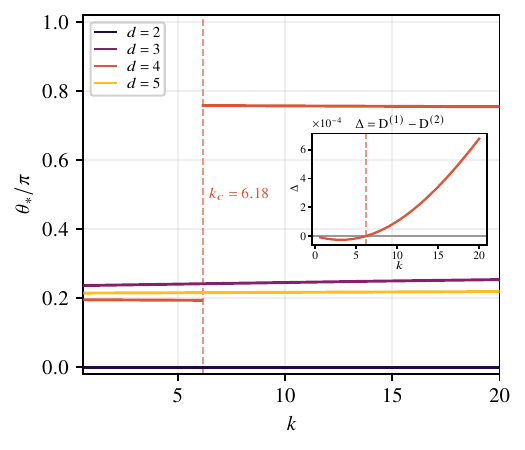}
    \caption{%
        Optimal drive angle $\theta_*(k)$ minimizing the
        \Renyi $k$-divergence $\mathrm{D}_k(P_0|| P_F)$ between the initial and
        final bath distributions, for bath qudit dimensions $d=2$–$5$ and parameters 
        $\beta=0.25$, $\omega_{\mathcal{S}}=1$, $\omega_B=0.3$, $\delta=1.5$.
        For $d=2$, the minimum is pinned at $\theta_*=0$, while for $d\geq 3$ it
        acquires $k$-dependence. At $d=4$, the drive supports two
        local minima at $\theta_*\approx0.19\pi$ and $0.76\pi$, and the
         optimal protocol switches discontinuously between them at
        $k_c\approx 6.18$ (dashed). Inset: the difference in \Renyi divergence
        $\Delta=\mathrm{D}^{(1)}-\mathrm{D}^{(2)}$ of the two minima crosses zero at
        $k_c$.
    }
    \label{fig:theta-transition}
\end{figure}

We now modify the setup, keeping the system $\mathcal{S}$ and $\mathcal{H}_{\mathcal{S}}$ identical, but promoting the bath $\mathcal{B}$ to be a $d$-state qudit with values $b\in \{0,\cdots, d-1\}$ and modifying its Hamiltonian to be $H_{\mathcal{B}} = \sum_{b=0}^{d-1} b\omega_{\mathcal{B}}\ketbra{b}{b}$. Bath measurements are still made in the computational basis. Accordingly, we modify the unitary drive to act as $U\ket{00}=\ket{00}$, $U\ket{1,d-1}=\ket{1,d-1}$, and 
\begin{equation}
    U \begin{pmatrix}
        \ket{0,b+1}\\
        \ket{1,b}
    \end{pmatrix}= R(\theta_b) \begin{pmatrix}
        \ket{0,b+1}\\
        \ket{1,b}
    \end{pmatrix}, b=0,\cdots, d-2,
\end{equation}
where $\theta_b=\theta -b\delta $. Here, $\theta$ remains a tunable control parameter and $\delta$ is some fixed offset so that the rotation angle $\theta_b$ in each two-state sector is set by a single parameter while also varying by sector. As before, the explicit form of $U$ is detailed in the Appendix. The optimal drive $\theta_*$  minimizes the \Renyi $k$-divergence, $d_\theta \mathrm{D}_k(P_0||P_F)=0$. Due to its analytic intractability for generic $d$, we find $\theta_*$ as a function of $k$ numerically for several $d$ in Fig.~\ref{fig:theta-transition}. For $d=2$, we recover the qubit result, with $\theta_*=0 \pmod{\pi}$ independent of $k$. For $d\geq 3$, the optimum acquires a weak $k$-dependence, and at $d=4$, we find a transition: at a critical value $k_c\approx 6.18$, the global minimum switches discontinuously between two competing, nearly degenerate minima $\theta_*^{(1)}$ and $\theta_*^{(2)}$. The inset displays this competition through the difference in divergence $\Delta\coloneq  \mathrm{D}_k(P_0||P_F(\theta_*^{(1)}))-\mathrm{D}_k(P_0||P_F(\theta_*^{(2)}))\equiv \mathrm{D}^{(1)}-\mathrm{D}^{(2)}$. For $k<k_c$, the two minima are nearly degenerate with $\theta_*^{(1)}$ the global minimum ($\Delta<0$), for $k=k_c$, the divergences cross ($\Delta =0$), and for $k>k_c$ the branch $\theta_*^{(2)}$ (which protects the low-probability tail events penalized at large $k$) becomes the global minimum ($\Delta >0$). Finally, in non-trivial cases ($\delta \neq 0$), the optimal solution demands $\theta \neq n\pi$, which imples that minimizing the disturbance to the bath \emph{strictly requires} entanglement generation between the system and the bath.

Our Renyi-$k$ minimization framework encapsulates and extends approaches of state preparation used in various fields of quantum science. For example, taking $k=1/2$ yields (for diagonal states) the quantum state fidelity $\mathrm{F}(\rho,\sigma) = \tr (\sqrt{\sqrt{\rho}\sigma\sqrt{\rho}})^2$, i.e. we have $\mathrm{D}_{1/2}(P_0||P_F)=-\log(\mathrm{F}(\rho_0^\mathcal{B},\rho_F^\mathcal{B}))$, where $\rho_0^\mathcal{B}$ and $\rho_F^\mathcal{B}$, are the bath marginals before and after the drive, respectively, which may be used as a cost function. When the distributions $P_F$ and $P_0$ cannot be made equal, the \Renyi-$k$ minimization allows tunable tail-event penalization by varying $k$.

\paragraph*{Discussion.}
In this Letter, we establish a fully quantum, arbitrary-bath Jarzynski equality that circumvents the destructive nature of standard work-measurement schemes. By leveraging the resource-theoretic framework, our \Renyi-Jarzynski equality places an exact thermodynamic constraint on the higher moments of non-equilibrium work distributions while allowing the system to evolve coherently under the drive. One might object to calling this expression ``quantum" since the \Renyi divergence only compares classical distributions and is blind to coherences in the bath state. We respond in two ways. First, the objection understates what the classical marginals see: the drive entangles $\mathcal{S}$ and $\mathcal{B}$, and this entanglement affects the classical marginal $P_F$. As we show below, minimizing the drift in $P_F$ can \textit{require} generating system-bath entanglement (see Fig.~\ref{fig:theta-transition}). Second, we grant that coherences are
discarded, but the \Renyi divergence still lower-bounds its quantum
counterpart. Among several inequivalent quantum
generalizations of the classical \Renyi divergence, we adopt the sandwiched divergence
\begin{equation}
\tilde{\mathrm{D}}_k(\rho||\sigma) \coloneq \frac{1}{k-1}\log\left(\tr\left(\sigma^{\frac{1-k}{2k}}\rho\sigma^{\frac{1-k}{2k}}\right)^k \right),    
\end{equation}
 as it satisfies the data processing inequality across all orders $k$ that we consider: $\tilde{\mathrm{D}}_{k}(\rho||\sigma) \geq \tilde{\mathrm{D}}_{k}(\mathcal{N}(\rho)||\mathcal{N}(\sigma))$ for any CPTP channel $\mathcal{N}$ and $k\geq 1/2$ \cite{Beigi_2013, Muller_Lennert_2013, Frank_2013}. Taking $\mathcal{N}$
to be the completely dephasing channel in the eigenbasis of $\mathcal{O}_\mathcal{B}$ (i.e. $\rho_{\mathcal{B}}\to \rho_{\mathcal{B}}^{\mathrm{diagonal}}$), and
noting that $\tilde{\mathrm{D}}_k$ reduces to the classical $\mathrm{D}_k$ on diagonal
states, we obtain
$\tilde{\mathrm{D}}_k(\rho_0^\mathcal{B}||\rho_F^\mathcal{B}) \geq
\mathrm{D}_k(P_0||P_F)$. The RJE therefore never overestimates the drift of the bath marginal. While we cannot infer from the work statistics alone how much the discarded coherences contribute,  any bath drift the RJE reports is guaranteed to be present in the full quantum state. The distributions $P_0$ and $P_F$ (and hence the divergence $\mathrm{D}_k(P_0||P_F)$) depend on the measured observable $\mathcal{O}_B$; optimizing over $\mathcal{O}_B$ provides the tightest lower bound on $\tilde{\mathrm{D}}_k$, which will generically remain strict unless $\rho_0^\mathcal{B}$ and $\rho_F^\mathcal{B}$ commute.

Beyond its implications for quantum thermodynamics, we demonstrate the practical utility of our RJE as a rigorous cost function for state preparation and gate implementation when preserving the state of a finite bath is crucial. Our toy model exhibits two features of interest: a transition in the global minimum as a function of \Renyi-order $k$ (which would be invisible to other cost functions) and a regime in which entanglement generation between the system and bath is required to minimize bath drift. Applying our RJE to more complex systems, such as noisy intermediate-scale quantum (NISQ) devices, to minimize cross-talk, presents a promising direction for protecting quantum states during non-adiabatic operations. 
As in the Sagawa-Ueda generalization of the Jarzynski equality~\cite{Sagawa_2010}, the information obtained would enter our fluctuation theorem directly; how such feedback modifies our $k$-copy fluctuation theorem is a natural extension for future work.  

Additionally, we believe Result~\ref{quasiclassicalrenyik} generalizes to purely statistical contexts whenever one wishes to hold an auxiliary random variable fixed. From a computational perspective, estimating \Renyi $k$-divergences is quite resource-intensive: Ref.~\cite{Renyidivergencebound} established that it takes $\Omega(D^{1/2})$ samples, where $D$ is the alphabet size. Result \ref{quasiclassicalrenyik} recasts the \Renyi divergence in terms of the expectation value of the exponentiated work over the $k$-fold projected ensemble. Because this estimator samples a work average instead of reconstructing $P_0$ and $P_F$, its cost is determined by the work variance, rather than the alphabet size $D$. Whether this yields practical advantages remains open: like the Jarzynski free energy estimator \cite{jarzysnkifreeestimator}, it is well-behaved when the work distribution is sharply peaked and the bath measurements lack heavy tails, but this near-equilibrium regime is also where the bath is barely perturbed and the \Renyi divergence is small. Understanding the intermediate regime where the estimator is both tractable and informative may open up promising new applications in Monte Carlo simulations and machine learning.

\begin{acknowledgments}
\paragraph*{Acknowledgments.}
We would like to acknowledge Guangkuo Liu, Charles Marrder, Katherine Slattery, and Amit Vikram for helpful discussions. Claude Opus 4.8 and Opus 5 assisted with debugging and vectorizing the Python code used to numerically solve the optimal drive angle $\theta_*$, and in formatting Fig.~\ref{fig:theta-transition}. The text and scientific results are all our own. 
\end{acknowledgments}

\nocite{*}
\bibliography{refs}% Produces the bibliography via BibTeX.
\clearpage % Flushes any floating figures/tables and starts a fresh page
\onecolumngrid % Forces the rest of the document into a single, wide column
\appendix

\section{Derivational details of results}

\subsection{Derivation of Result \ref{res:gibbsstochasticfine}}\label{appendixA}
%This appendix details the derivation of Result \ref{res:gibbsstochasticfine}. To maintain generality throughout the intermediate steps, we introduce a generic bath operator $E_b^\mathcal{B}$, which we will eventually substitute with our specific measurement operator, $\frac{\Pi_b^\mathcal{B}}{\sqrt{P_F(b)}}$. We begin by expanding our definition:
For generality, we work with bath POVMs
$E_b^\mathcal{B}$, and eventually substitute normalized rank-1 measurements $\Pi_b^\mathcal{B}\big /\sqrt{P_F(b)}$ when needed. By definition~\eqref{eq:gibbschannel}, we have 
\begin{align}
    \Omega_b\left(\frac{\ee^{-\beta H_\mathcal{S}}}{Z_\mathcal{S}}\otimes\rho_\mathcal{W}\right)\coloneq &\text{Tr}_\mathcal{W}\left(\mathcal{J}_{H_\mathcal{S}'+H_\mathcal{W}}\Phi_b\mathcal{J}_{-H_\mathcal{W}}\left(\frac{\ee^{-\beta H_\mathcal{S}}}{Z_\mathcal{S}}\otimes \rho_\mathcal{W}\right)\right)  \notag \\ 
    =&\frac{1}{Z_\mathcal{B}Z_\mathcal{S}}\text{Tr}_\mathcal{W}\left(\mathcal{J}_{H_\mathcal{S}'+H_\mathcal{W}}\left( \text{Tr}_\mathcal{B} \left(E_b^{\mathcal{B}} U \mathcal{J}_{-(H_\mathcal{S}+H_\mathcal{W}+H_\mathcal{B})}(\mathbb{I}_{\mathcal{S}\mathcal{B}}\otimes \rho_\mathcal{W})U^{\dagger}(E_b^{\mathcal{B}})^{\dagger}\right)  \right)\right).
\end{align}
where we used the definition of $\Phi_b$ and that $\mathcal{J}_{-H_\mathcal{W}}(\ee^{-\beta H_\mathcal{S}}\otimes \mathbb{I}_\mathcal{B}\otimes \rho_\mathcal{W}) \ee^{-\beta H_\mathcal{B}}= \mathcal{J}_{-(H_\mathcal{S}+H_\mathcal{W}+H_\mathcal{B})}( \mathbb{I}_{\mathcal{S}\mathcal{B}}\otimes \rho_\mathcal{W})$ holds.
Next, we utilize energy conservation~\eqref{eq:energy_cons} %$U(H_\mathcal{S}+H_\mathcal{W}+H_\mathcal{B})U^{\dagger} = H_{\mathcal{S}'}+H_\mathcal{W}+H_\mathcal{B}$ 
to rewrite the argument of $\mathcal{J}_{-(H_\mathcal{S}+H_\mathcal{W}+H_\mathcal{B})}$ as
\begin{align}
  \Omega_b\left(\frac{\ee^{-\beta H_\mathcal{S}}}{Z_\mathcal{S}}\otimes\rho_\mathcal{W}\right)=\frac{1}{Z_\mathcal{B}Z_\mathcal{S}}  \text{Tr}_W\left(\mathcal{J}_{H_\mathcal{S}'+H_\mathcal{W}}\left(\text{Tr}_\mathcal{B} \left(E_b^\mathcal{B}  \mathcal{J}_{-(H_\mathcal{S}'+H_\mathcal{W}+H_\mathcal{B})}(U\mathbb{I}_{\mathcal{S}\mathcal{B}}\otimes \rho_\mathcal{W}U^{\dagger})(E_b^\mathcal{B})^{\dagger} \right)  \right)\right)\, .
   \end{align}
Because the generic bath operator $E_b^\mathcal{B}$ commutes with the system and work Hamiltonians $H_{\mathcal{S}}'+H_{\mathcal{W}}$, we can bring the $\mathcal{J}_{H_\mathcal{S}'+H_\mathcal{W}}$ inside the partial trace over the bath. %This allows us to cancel the corresponding s
System and work components of $\mathcal{J}$ channels cancel, simplifying the expression to 
\begin{align}
\Omega_b\left(\frac{\ee^{-\beta H_\mathcal{S}}}{Z_\mathcal{S}}\otimes\rho_\mathcal{W}\right)=\frac{1}{Z_\mathcal{S}Z_\mathcal{B}} \text{Tr}_{\mathcal{B}\mathcal{W}}\left(E_b^\mathcal{B}  \mathcal{J}_{-H_\mathcal{B}}(U\mathbb{I}_{\mathcal{S}\mathcal{B}}\otimes \rho_\mathcal{W}U^{\dagger})(E_b^\mathcal{B})^{\dagger} \right) \, .
\end{align}
%Now plugging in our 
By definition of $\mathcal{J}_{H_\mathcal{B}}$ and using the cyclicity of the trace we obtain 
\begin{align}
   \Omega_b\left(\frac{\ee^{-\beta H_\mathcal{S}}}{Z_\mathcal{S}}\otimes\rho_\mathcal{W}\right) =&\frac{1}{Z_\mathcal{S}Z_\mathcal{B}}\text{Tr}_{\mathcal{B}\mathcal{W}} \left(E_b^\mathcal{B} \ee^{-\beta H_\mathcal{B}/2}(U\mathbb{I}_{\mathcal{S}\mathcal{B}}\otimes \rho_\mathcal{W}U^{\dagger})e^{-\beta H_\mathcal{B}/2} )(E_b^\mathcal{B})^{\dagger} \right) \notag  
   \\=
   &\frac{1}{Z_\mathcal{S}Z_\mathcal{B}}\text{Tr}_{\mathcal{B}\mathcal{W}} \left(\ee^{-\beta H_\mathcal{B}/2}(E_b^\mathcal{B})^{\dagger}E_b^\mathcal{B} \ee^{-\beta H_\mathcal{B}/2}(U\mathbb{I}_{\mathcal{S}\mathcal{B}}\otimes \rho_\mathcal{W}U^{\dagger}) \right) \, .
\end{align}
Finally, Result 1 of Ref.~\cite{Masanes_2017} (which states that tracing over the ideal weight yields $\text{Tr}_\mathcal{W}(U (\mathbb{I}_{\mathcal{S}\mathcal{B}}\otimes \rho_\mathcal{W})U^{\dagger})=\mathbb{I}_{\mathcal{S}\mathcal{B}}$, provided $U$ possesses translational invariance in the weight $\mathcal{W}$) and taking $E_b^\mathcal{B} = {\Pi_b^\mathcal{B}}\big/{\sqrt{P_F(b)}}$ implies our Result~\ref{res:gibbsstochasticfine}, i.e.,
\begin{align}
   \Omega_b\left(\frac{\ee^{-\beta H_\mathcal{S}}}{Z_\mathcal{S}}\otimes\rho_\mathcal{W}\right)=\frac{1}{Z_\mathcal{S}Z_\mathcal{B}P_F(b)}\text{Tr}_{\mathcal{B}} \left(\ee^{-\beta H_\mathcal{B}/2}\Pi_b^\mathcal{B} \ee^{-\beta H_\mathcal{B}/2} \mathbb{I}_{\mathcal{S}\mathcal{B}} \right)=\frac{ \text{Tr}_{\mathcal{B}}\left(\ee^{-\beta H_\mathcal{B}}\Pi_b^\mathcal{B} \right) }{Z_\mathcal{S}Z_\mathcal{B}P_F(b)}\mathbb{I}_{\mathcal{S}} 
 \equiv \frac{P_0(b)}{Z_\mathcal{S}P_F(b)} \mathbb{I}_{\mathcal{S}}.
\end{align}
Here, $P_0(b)=\text{Tr}_\mathcal{B}\left( \Pi_b^{\mathcal{B}} {\ee^{-\beta H_\mathcal{B}}}\right)\big/{Z_\mathcal{B}} $ is the probability of obtaining measurement outcome $b$ if the bath were completely isolated. 

\subsection{Derivation of Result~\ref{res:quasiclassicalmeasurementjarz}}

Recall the action $\Phi_b(\ketbra{s}{s}  \otimes \ketbra{0}{0}_\mathcal{W} ) = \sum_{s',w} P(s',w|s,b)\ketbra{s'}{s'}\otimes \ketbra{w}{w}$ of $\Phi_b$~\eqref{eq:CP_phi} on a joint system-work state and note that we set $\rho_\mathcal{W}=\ketbra{0}{0}_{\mathcal{W}}$ without loss of generality.

Let us further denote the spectrum of $H_{\mathcal{S}(')}$ as $E_{s(')}$, in terms of which we expand the channel $\Phi_b$ as 
\begin{align}\label{eq:Yb_start}
    \mathcal{Y}_b\left(\frac{\ee^{-\beta H_\mathcal{S}}}{Z_\mathcal{S}}\otimes |0\rangle\langle0|_\mathcal{W}\right) &= \frac{1}{Z_\mathcal{S}}\text{Tr}_{\mathcal{S}\mathcal{W}}\left(\mathcal{J}_{H_\mathcal{W}}\Phi_b(\ee^{-\beta H_\mathcal{S}}\otimes \ee^{-\beta H_\mathcal{W}}|0\rangle\langle0|_\mathcal{W})\right) = \frac{1}{Z_\mathcal{S}}\sum_{s,s',w}e^{-\beta (E_s-w)}P(s',w |s,b) \, .
\end{align}
To make the notation suggestive, we rewrite~\eqref{eq:Yb_start} in terms of the initial free energy `density' $f_s \equiv E_s +\log P_0(s)/\beta$ of state $s$, where $e^{-\beta (E_s-w)}P(s',w |s,b)  = \ee^{-\beta \left(f_s-w\right)}P(s',w |s,b)P_0(s)$.  
The remaining sum over $s'$ may be trivially performed, and we may massage the expression using $P(w|s,b)P_0(s)=P(w,s|b)$ to   
\begin{align}\label{eq:Yb_intermediate}
    \mathcal{Y}_b\left(\frac{\ee^{-\beta H_\mathcal{S}}}{Z_\mathcal{S}}\otimes |0\rangle\langle0|_\mathcal{W}\right) &= \frac{1}{Z_{\mathcal{S}}}\sum_{s,w}e^{-\beta \left(f_s-w\right)}P(w, s|b) \equiv\frac{1}{Z_\mathcal{S}} \mathbb{E}_{P(s,w|b)}[ \ee^{-\beta \left(f_s-w\right)}] \, .
\end{align}
If we assume $P_0(s)= {\ee^{-\beta E_s}}\big/{Z_\mathcal{S}}$, i.e., that the system initially starts in a thermal state, the fine-grained free energy simplifies to a constant $f_s=-\log(Z_\mathcal{S})/\beta$ which is entirely independent of the specific state $s$.
This gets rid of the average over system states $\mathbb{E}_{s}$ in~\eqref{eq:Yb_intermediate}, yielding Result~\ref{res:quasiclassicalmeasurementjarz}, which reads 
\begin{equation}
  \mathbb{E}[ \ee^{\beta w}]_{ P_F(w|b)} =  \frac{P_0(b) Z_{\mathcal{S}}'}{P_F(b) Z_\mathcal{S}}. 
\end{equation}

\section{Details on toy examples}

\subsection{Derivation of Simple Toy Qubit} \label{ToyQubit1}

Suppose our system and bath are each a single qubit, respectively, with the Hamiltonian
\begin{equation}
    H_{\mathcal{SBW}}= \omega_\mathcal{S}\ketbra{1}{1}_\mathcal{S}+\omega_\mathcal{B}\ketbra{1}{1}_\mathcal{B}+x_\mathcal{W} \equiv  H_\mathcal{S}+H_\mathcal{B}+H_\mathcal{W}\, , 
\end{equation}
and we want to change the Hamiltonian such that the system frequency changes from $\omega_\mathcal{S}\to \omega_\mathcal{S}'$. For brevity, we define $H\equiv H_\mathcal{S}+H_\mathcal{B}+H_\mathcal{W}$ and $H'\equiv H_\mathcal{S}'+H_\mathcal{B}+H_\mathcal{W}$.
The initial states of interest are Gibbs states $\rho_\mathcal{SB}$ on $\mathcal{SB}$ and with the weight system in the pure zero state $\ket{0}\bra{0}$. Expanding in the computational basis $\{\ket{0}, \ket{1}\}$ (i.e., the $\sigma_z$ eigenbasis on each subsystem), we have 
\begin{align}
    \rho_{\mathcal{SBW}} &= \frac{1}{Z_\mathcal{S}Z_\mathcal{B}} \ee^{-\beta \omega_\mathcal{S}\ketbra{1}{1}_\mathcal{S}}\otimes \ee^{-\beta \omega_\mathcal{B}\ketbra{1}{1}_\mathcal{B}} \otimes \ketbra{0}{0}_\mathcal{W}\, \notag  \\
    &=\frac{1}{Z_\mathcal{S}Z_\mathcal{B}}\left(\ketbra{00}{00}+\ketbra{10}{10}e^{-\beta\omega_\mathcal{S}}+\ketbra{01}{01}e^{-\beta\omega_\mathcal{B}}+e^{-\beta (\omega_\mathcal{S}+\omega_\mathcal{B})}\ketbra{11}{11}\right)  \otimes \ketbra{0}{0}_\mathcal{W} \, \notag \\
    &\equiv \left(P_{00}\ketbra{00}{00}+P_{10}\ketbra{10}{10}+P_{01}\ketbra{01}{01}+P_{11}\ketbra{11}{11}\right)  \otimes \ketbra{0}{0}_\mathcal{W}\, , 
\end{align}
where we denote $\ket{a}_\mathcal{S}\otimes \ket{b}_\mathcal{B}\equiv \ket{ab} $ and we call $P_{ab}$ the joint probability of $(s=a,b=b)$. The Hamiltonian takes the form 
\begin{equation}
    H_\mathcal{S}+H_\mathcal{B} = \omega_\mathcal{S}\ketbra{1}{1}_\mathcal{S} \otimes \mathbb{I}_{\mathcal{B}}+ \omega_\mathcal{B} \mathbb{I}_{\mathcal{S}}\otimes \ketbra{1}{1}_\mathcal{B}= (\omega_\mathcal{S}+\omega_\mathcal{B})\ketbra{11}{11}+\omega_\mathcal{S}\ketbra{10}{10}+ \omega_\mathcal{B}\ketbra{01}{01}\, .
\end{equation}
We model the driving process such that our qubits can be jointly relaxed/excited without any entanglement, but will develop some entanglement if they do not start in the same state. Mathematically, this is encoded in the parity of the basis states of $\mathcal{SB}$, where we keep the even sector $\ket{00}, \ket{11} \to \ket{00}, \ket{11}$ while rotating the odd sector as $(\ket{01}, \ket{10}) \to (\ket{01}, \ket{10})R(\theta)^T  = (\cos(\theta)|01\rangle -\sin(\theta)|10\rangle,   \cos(\theta)\ket{10} +\sin(\theta)\ket{01}$
where $\theta$ is an angle parametrizing the entanglement, such that the nonentangling limit is $\theta = 0$ (with $R(0) = \mathbb{I}$). With the action on $\mathcal{SB}$ fixed, the last ingredient to find $U$ is to fix what happens to $\mathcal{W}$ during the thermodynamic process.  
Translational invariance in $\mathcal{W}$ implies that we may write $U=\sum_w A_{\mathcal{SB}}^w\otimes \ee^{-iw\hat{p}}\equiv \sum_w A_{\mathcal{SB}}^w\otimes T_{w}$ where $T_w$ is a translation operator on $\mathcal{W}$ that keeps track of the work due to the change in $H$. 
For the particular process considered (rotation in the odd-parity sector), it reads 
\begin{multline}
    U  =\ketbra{00}{00}\otimes T_{0}+ \ketbra{11}{11}\otimes T_{\omega_\mathcal{S}-\omega_\mathcal{S}'} +  \cos(\theta)\ketbra{10}{10}\otimes T_{\omega_\mathcal{S}-\omega_\mathcal{S}'} \\
    +\cos(\theta)\ketbra{01}{01} \otimes T_{0}+\sin(\theta) \ketbra{01}{10}\otimes T_{\omega_\mathcal{S}-\omega_\mathcal{B}} -\sin(\theta)\ketbra{10}{01}\otimes T_{\omega_{\mathcal{B}}-\omega_\mathcal{S}'}\, .
\end{multline}
For this unitary to generate a thermodynamic process, we must have $UH=H'U$, which we verify by direct computation on the energy eigenbasis.

Straightforward manipulations yield the action of $UH$ on the eigenvectors, which read
\begin{align}\label{eq:U_on_1}
    U H\ket{00} \otimes \ket{E}  &= E \ket{00} \otimes \ket{E} \, , \\
    UH \ket{11} \otimes \ket{E} &= (\omega_\mathcal{S}+\omega_\mathcal{B}+E) \ket{11}\otimes \ket{E+\omega_\mathcal{S}-\omega_\mathcal{S}'} \, , \\
    U H\ket{01} \otimes \ket{E}     &=(\omega_\mathcal{B}+E)(\cos(\theta)\ket{01}\otimes\ket{E} -\sin(\theta)\ket{10}\otimes \ket{E+\omega_\mathcal{B}-\omega_\mathcal{S}'} )\, , \\
    UH\ket{10} \otimes \ket{E} &= (\omega_\mathcal{S}+E)(\cos(\theta)\ket{10}\otimes \ket{E+\omega_\mathcal{S}-\omega_\mathcal{S}'}+\sin(\theta)\ket{01}\otimes \ket{E+\omega_\mathcal{S}-\omega_\mathcal{B}} )\, .
\end{align}

The opposite direction $H'U$ yields the same kets, concluding $UH = H'U$.

To find the optimal $\theta$ to minimize entanglement, we further need the evolution of the density matrix on $\mathcal{SBW}$, which reads
\begin{align*}
    U\rho_{\mathcal{SBW}} U^{\dagger}&= P_{00}\ketbra{00}{00}\otimes \ketbra{0}{0} +P_{11}\ketbra{11}{11} \otimes \ketbra{\omega_\mathcal{S}-\omega_\mathcal{S}'}{ \omega_\mathcal{S}-\omega_\mathcal{S}'}\\
    &+P_{10}\cos^2(\theta)\ketbra{10}{ 10} \otimes \ketbra{\omega_{\mathcal{S}}-\omega_\mathcal{S}'}{  \omega_{\mathcal{S}} -\omega_\mathcal{S}'}+P_{01}\cos^2(\theta)\ketbra{01}{01}\otimes \ketbra{0}{0} \\
    &+P_{10}\sin^2(\theta)\ketbra{01}{ 01} \otimes \ketbra{\omega_\mathcal{S}-\omega_\mathcal{B}}{ \omega_\mathcal{S}-\omega_\mathcal{B}}+P_{01}\sin^2(\theta)\ketbra{10}{10 }\otimes \ketbra{\omega_\mathcal{B}-\omega_\mathcal{S}'}{ \omega_\mathcal{B}-\omega_\mathcal{S}'} \\
    &+P_{10}\cos(\theta)\sin(\theta)\ketbra{10}{ 01}\otimes \ketbra{\omega_\mathcal{S}-\omega_\mathcal{S}'}{ \omega_\mathcal{S}-\omega_\mathcal{B}}  - P_{01}\cos(\theta)\sin(\theta) \ketbra{10}{ 01} \otimes \ketbra{ \omega_\mathcal{B}-\omega_\mathcal{S}'}{ 0}
\\&+P_{01}\cos(\theta)\sin(\theta)\ketbra{01}{ 10}\otimes \ketbra{\omega_\mathcal{S}-\omega_\mathcal{B}}{\omega_\mathcal{S}-\omega_\mathcal{S}'}-P_{10}\cos(\theta)\sin(\theta)\ketbra{01}{ 10} \otimes \ketbra{0}{ \omega_\mathcal{B}- \omega_\mathcal{S}'}\, .
\end{align*}

The $b$ marginals can be read off, which are 
\begin{equation}
    P_F(b=0) =  P_{00}+P_{01}\sin^2(\theta) +P_{10}\cos^2(\theta)\, , \quad 
    P_F(b=1)=P_{11}+ P_{01}\cos^2(\theta)+P_{10}\sin^2(\theta)\, .
\end{equation}
To find when $\mathrm{D}_k(P_0||P_F)$ is minimized we note that $ P_F(b=0;\theta=n\pi)=P_0(b=0) $ and $P_F(b=1;\theta=n\pi)=P(b=1)$, which means $\mathrm{D}_k(P_0||P_F)=0$ for
$\theta= n\pi$ with integer $n$.

\subsection{Toy Qudit With Nontrivial Entanglement}\label{app:ToyQubit3}
We keep the system Hamiltonian the same but we modify the bath Hamiltonian to comprise a $d$-state qudit with an increasing frequency: $H_{\mathcal{B}} = \sum_{b=0}^{d-1} b\omega_{\mathcal{B}}\ketbra{b}{b}$. The initial bath marginal is given by $P_0(b) =\exp(-\beta b\omega_{\mathcal{B}})/Z_{\mathcal{B}} $. As before, we will denote $P_{sb}$ to be the joint probabilities for $\mathcal{SB}$ to be measured in $sb$ in the energy eigenbasis. We build $U$ as a chain of nearest state swaps on the bath for $b=0,\cdots, d-2$, while fixing identical boundaries
\begin{align}
\ket{00} &\to \ket{00}, \nonumber \quad 
&&\ket{1,d-1} \to \ket{1,d-1},  \nonumber\\
\ket{0,b+1} &\to \cos(\theta_b)\ket{0,b+1} - \sin(\theta_b)\ket{1,b}, \nonumber 
&&\ket{1,b} \to \cos(\theta_b)\ket{1,b} + \sin(\theta_b)\ket{0,b+1},
\end{align}
where $\theta_b =\theta-b\delta $, for some fixed hardware offset $\delta$. Because $\theta=0$ will generate entanglement depending on the phase $\delta$, the optimal choice of $\theta$ is no longer trivial. Like before, the unitary $U$ on $\mathcal{SBW}$ is given by tensoring the $\mathcal{SB}$ transitions with the corresponding translation operators on the weight, i.e.,
\begin{align}
    U= \ketbra{00}{00} +\sum_{b=0}^{d-2}H(\theta_b;b) + \ketbra{1, d-1}{1,d-1}\otimes T_{\omega_{\mathcal{S}}-\omega_{\mathcal{S}}'},
\end{align}
where we have dropped tensor products with $T_0$ since $T_0=\mathbb{I}$ and 
\begin{align}
    H(\theta_b;b) &\coloneq \cos(\theta_b )\ketbra{0,b+1}{0,b+1}+\cos(\theta_b) \ketbra{1,b}{1,b} \otimes T_{\omega_{\mathcal{S}}-\omega_{\mathcal{S}}'}\\
    &+\sin(\theta_b)\ketbra{0,b+1}{1,b} \otimes T_{\omega_{\mathcal{S}}-\omega_\mathcal{B}} -\sin(\theta_b)\ketbra{1,b}{0,b+1}\otimes T_{\omega_B-\omega_{\mathcal{S}}'}.
\end{align}
To verify that this unitary properly updates our Hamiltonian, we show $UH=H'U$ by applying both sides of the equation to the basis state $\ket{sb}\otimes \ket{E}$,
\begin{align}
    UH\ket{00}\otimes \ket{E} &= E\ket{00}\otimes \ket{E},  \\
    UH\ket{1,d-1}\otimes \ket{E} &=(E+(d-1)\omega_{\mathcal{B}}+\omega_{\mathcal{S}})\ket{1,d-1}\otimes \ket{E+\omega_{\mathcal{S}}-\omega_{\mathcal{S}}'},\\
    UH\ket{0,b+1}\otimes \ket{E}&= (E+(b+1)\omega_{\mathcal{B}})(\cos(\theta_b)\ket{0,b+1}\otimes \ket{E} -\sin(\theta_b)\ket{1,b}\otimes \ket{E+\omega_{\mathcal{B}}-\omega_{\mathcal{S}}'}),\\
    UH\ket{1,b}\otimes \ket{E} &= (E+\omega_{\mathcal{S}}+b\omega_{\mathcal{B}})(\cos(\theta_b) \ket{1,b}\otimes \ket{E+\omega_{\mathcal{S}}-\omega_{\mathcal{S}}'}+ \sin(\theta_b)\ket{0,b+1}\otimes \ket{E+\omega_{\mathcal{S}}-\omega_{\mathcal{B}}})\, .
\end{align}
The opposite direction $H'U$ yields identical states, showing that $UH=H'U$. Next, we must again compute the final state $\rho'_{\mathcal{SBW}}= U\rho_{\mathcal{SBW}}U^{\dagger}$ after this unitary drive. Since we only care about the bath marginals, we shall drop the coherence terms and trace over $\mathcal{W}$:
\begin{align}
   \tr_{\mathcal{W}} \left(U\rho_{\mathcal{SBW}}U^{\dagger}\right)&= \sum_{b=0}^{d-2} (P_{0,b+1} \cos^2(\theta_b)+P_{1,b}\sin^2(\theta_b))\ketbra{0,b+1}{0,b+1} +(P_{0,b+1}\sin^2(\theta_b)+P_{1,b}\cos^2(\theta_b))\ketbra{1,b}{1,b}\\
   &+P_{0,0}\ketbra{00}{00}+ P_{1,d-1}\ketbra{1,d-1}{1,d-1}+\text{coherences}
\end{align}
The bath marginals can thus be read off from the above by taking the partial trace in $\mathcal{S}$:
\begin{equation}
     P_F(b)=P_0(b) +c_{b-1}\sin^2(\theta_{b-1}) - c_{b}\sin^2(\theta_b),
\end{equation}
where we defined $c_{b}\coloneq  \exp(-\beta b\omega_{\mathcal{B}} )(\exp(-\beta \omega_{\mathcal{S}} )-\exp(-\beta \omega_{\mathcal{B}}))/Z_{\mathcal{S}}Z_{\mathcal{B}}$ and we let $c_{-1}=c_{d-1}=0$. Our \Renyi $k$-divergence between the initial and final bath marginals will be:
\begin{equation}
    \mathrm{D}_k(P_0||P_F) =\frac{1}{k-1}\log\left(\sum_{b=0}^{d-1} \frac{P_0(b)^k}{(P_0(b) +c_{b-1}\sin^2(\theta_{b-1}) - c_{b}\sin^2(\theta_b))^{k-1}} \right)
\end{equation}
We can minimize this by taking a derivative with respect to $\theta$ and setting it equal to zero. This will yield;
\begin{equation}
    \sum_{b=0}^{d-1} \left(\frac{P_0(b)}{P_F(b)} \right)^k\frac{dP_F(b)}{d\theta}=0
\end{equation}
Given its analytic intractability, we solve this expression numerically. For each $k$, we enumerate all local minima of $\mathrm{D}_k(P_0||P_F)$ over the period $\theta \in [0,\pi)$ and track each local minimum as a continuous branch in $k$. For the values of $d$ considered, we found at most two branches. The transition at $d=4$ is the point $k_c$ where the global minimum passes from one branch to the other. We confirm that this is a genuine crossing rather than a numerical artifact: the two minima are distinct and $k$-dependent, and their difference in divergence $\Delta\coloneq  \mathrm{D}_k(P_0||P_F(\theta_*^{(1)}))-\mathrm{D}_k(P_0||P_F(\theta_*^{(2)}))$ changes sign across $k_c$.

\end{document}